%% file: Rc_draft.tex
\documentclass[dvipsnames,aps,prl,twocolumn,showpacs,amsmath,amssymb,noeprint,nolongbibliography,nobibnotes]{revtex4-2}

\usepackage[dvipsnames]{xcolor}
\usepackage{graphicx}
\usepackage{dcolumn}
\usepackage{bm}

\definecolor{Indigo}{RGB}{51, 51, 153}
\usepackage[colorlinks,linkcolor=blue,citecolor=blue,urlcolor=Indigo]{hyperref}

\usepackage{url}
\usepackage{ulem}
\usepackage{amsmath}
\usepackage{amsfonts}
\usepackage{textcomp}
\usepackage{subfigure}
\usepackage{verbatim}
\usepackage{rotating}
\usepackage{longtable}
\usepackage{float}
\usepackage{ulem}
\usepackage{multirow} 
\usepackage{booktabs}
\usepackage{lineno}
\usepackage{orcidlink}
\usepackage{overpic}
\usepackage{tikz}
\usetikzlibrary{calc}

\newcommand{\BESIIIorcid}[1]{\orcidlink{#1}}

\begin{document}

\title{Precise Measurement of the Inclusive Cross Sections of Open-Charm  Mesons at Energies from 3.80 to 4.95 GeV}

\author{
M. Ablikim \textit{et al.}
}
\thanks{Full author list given at the end of the Letter}

\collaboration{BESIII Collaboration}


\begin{abstract}
Using $e^+e^-$ collision data  collected by the BESIII detector at center-of-mass energies from $3.80$ to $4.95~\text{GeV}$, the inclusive production cross sections of open-charm mesons, $\sigma(e^{+}e^{-}\to D^{0}/D^{+}/D_{s}^{+}+X)$, are measured by tagging a $D^0$, $D^+$, or  $D_s^+$ meson, where $X$ denotes any accompanying particles. Open-charm production dominates the total hadronic yield above threshold. Therefore these measurements provide a vital reference for understanding hadron production and for validating global $R$ value results. The results offer crucial inputs for investigating heavier charmonia and exotic states coupled to open-charm final states, while serving as a benchmark for comparison with the sum of measured exclusive channels.  This comparison directly evaluates the completeness of current exclusive datasets and clarifies the underlying hadronization mechanisms in the open-charm region.

\end{abstract}

  

\maketitle


The total hadronic cross section above the open-charm threshold in $e^+e^-$ annihilation, commonly expressed through the $R$ value, exhibits pronounced deviations from smooth perturbative-QCD (pQCD) expectations due to threshold effects, non-perturbative dynamics, and the presence of vector charmonium states such as $\psi(3770)$, $\psi(4040)$, $\psi(4160)$, and $\psi(4415)$~\cite{ablikim2008determination}. Historical measurements  of $R$ value by BES~\cite{BES:1999wbx,BES:2001ckj}, Mark~I~\cite{Siegrist:1981zp}, Crystal Ball~\cite{Osterheld:1986hw}, and PLUTO~\cite{Criegee:1981qx} suffer from limited energy coverage or precision, leaving the detailed line shapes above the threshold not fully established. Recent observations of  exotic charmonium-like structures, including the $Y(4230)$, $Y(4360)$, $Y(4660)$, and structures near $4.7~\mathrm{GeV}$~\cite{BESIII:2023wsc,BESIII:2023wqy,BESIII:2022kcv} further highlight the need for a precise characterization of hadronic production in this region. While these states are primarily observed in exclusive hidden-charm channels~\cite{BESIII:2016bnd,BESIII:2022kcv,BESIII:2023wqy}, open-charm meson production constitutes the dominant component of hadron production in this energy region~\cite{Wang:2025dur}.

A precise measurement of the inclusive production of open-charm mesons ($D^0$, $D^+$, and $D_s^+$) across a wide energy range, quantified via the $R_D$ value, thus provides a direct probe of the dominant hadronic processes above the threshold.  Compared with the $R$ measurements mentioned above, determining $R_D$ via tagged open-charm mesons offers distinct experimental advantages, including clean event topologies and strongly suppressed continuum light-hadron backgrounds.
Such a measurement provides a direct determination of the dominant open-charm contribution to hadron production, valuable input for the charm-quark mass determination~\cite{Kuhn:2007vp}, and important constraints on heavier charmonium and exotic charmonium-like states coupled to open-charm final states. Extensive exclusive open-charm measurements have been reported by BESIII~\cite{BESIII:2015tix, BESIII:2018iea, BESIII:2020gqt, BESIII:2020qkh, BESIII:2021yvc, BESIII:2022quc, BESIII:2023wsc, BESIII:2024ths, BESIII:2024zdh}, Belle~\cite{Belle:2006hvs,Belle:2007qxm,Belle:2007xvy,Belle:2009dus} and BaBar~\cite{BaBar:2006qlj,BaBar:2009elc,BaBar:2010plp} via direct $e^+e^-$ collisions and initial-state radiation (ISR), and comparisons between inclusive open-charm production and the sum of measured exclusive channels further provide a benchmark for assessing the completeness of current open-charm measurements, with important implications for $K$-matrix-based coupled-channel analyses of the vector charmonium spectrum~\cite{Uglov:2016orr,Husken:2022yik,Nakamura:2023obk,Husken:2024hmi}.
However, previous inclusive open-charm measurements by BES~\cite{BES:1999vjc} and CLEO-c~\cite{CLEO:2008ojp} were  restricted to  limited energy points and could not resolve the detailed energy dependence of this dominant contribution.

In this Letter, we present a measurement of the inclusive open-charm meson processes \( \sigma(e^+e^- \to D^0/D^+/D_s^+ + X) \), where \( X \) denotes any accompanying final-state particles, using data samples collected by the BESIII detector at 150 center-of-mass (c.m.) energy ($\sqrt{s}$)  points from 3.80 to \( 4.95~\text{GeV} \) \cite{BESIII:2015zbz,BESIII:2017lkp,BESIII:2020eyu,BESIII:2022ulv,BESIII:2022dxl}. Throughout this Letter, charge-conjugate modes are implicitly included.

The details of the design and performance of BESIII detector can be found in  Refs.~\cite{BESIII:2009fln,Huang:2022wuo}. Based on this setup, the inclusive open-charm meson cross sections  $\sigma(e^{+}e^{-}\to D^{0}X)$, $\sigma(e^{+}e^{-}\to D^{+}X)$, and $\sigma(e^{+}e^{-}\to D^{+}_{s}X)$ are measured using three selected decay modes: $D^{0}\to K^{-}\pi^{+}\pi^{+}\pi^{-}$, $D^{+}\to K^{-}\pi^{+}\pi^{+}$, and $D^{+}_{s} \to K^{-}K^{+}\pi^{+}$. The signal Monte Carlo (MC) samples, referred to as \textsc{hybrid} MC,  are generated using the \textsc{hybrid} event generator \cite{BESIII:2021wib,Ping:2016pms}, which combines the \textsc{conexc} \cite{Ping:2013jka}, \textsc{phokhara} \cite{Czyz:2008kw}, and \textsc{luarlw} models \cite{Andersson:1997xwk,Andersson:1999ui}. The \textsc{conexc} generator simulates a series of known exclusive open-charm processes, while the unknown processes containing one of the three open-charm mesons are modeled with \textsc{luarlw}. To prevent \textsc{luarlw} reproducing exclusive processes that are already generated with \textsc{conexc}, a veto procedure is implemented within the \textsc{hybrid} framework.  The phenomenological parameters of the {\sc luarlw}  component are determined through a dedicated tuning procedure  based on the inclusive hadronic events~\cite{BESIII:2021wib}. The decays $D^{0}\to K^{-}\pi^{+}\pi^{+}\pi^{-}$, $D^{+}\to K^{-}\pi^{+}\pi^{+}$, and $D^{+}_{s}\to K^{-}K^{+}\pi^{+}$ are simulated  according to the corresponding amplitude models~\cite{BESIII:2017jyh,CLEO:2008jus,BESIII:2020ctr}.

\begin{figure*}[htp]
    \centering
    \includegraphics[width=0.32\textwidth]{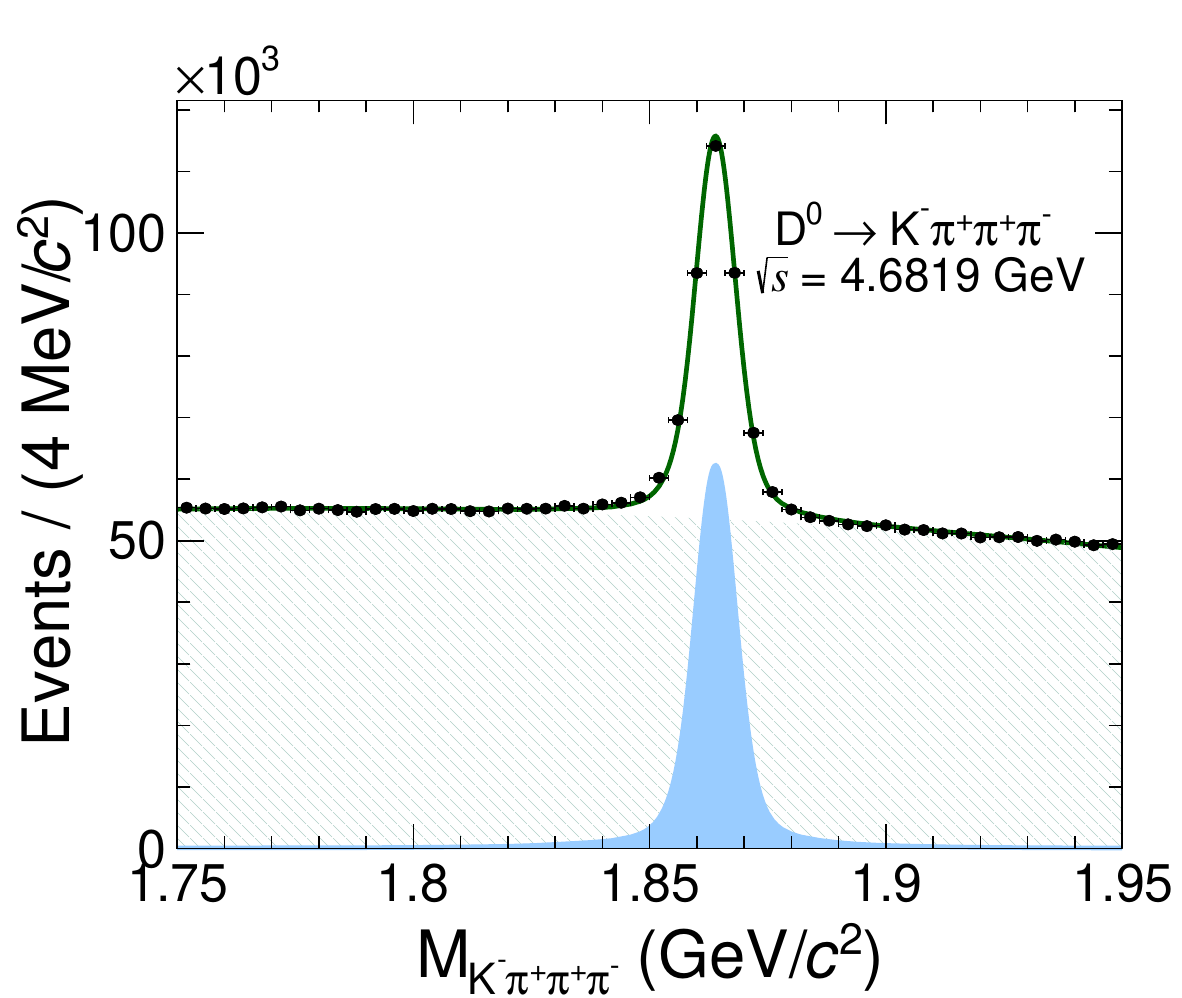}
    \includegraphics[width=0.32\textwidth]{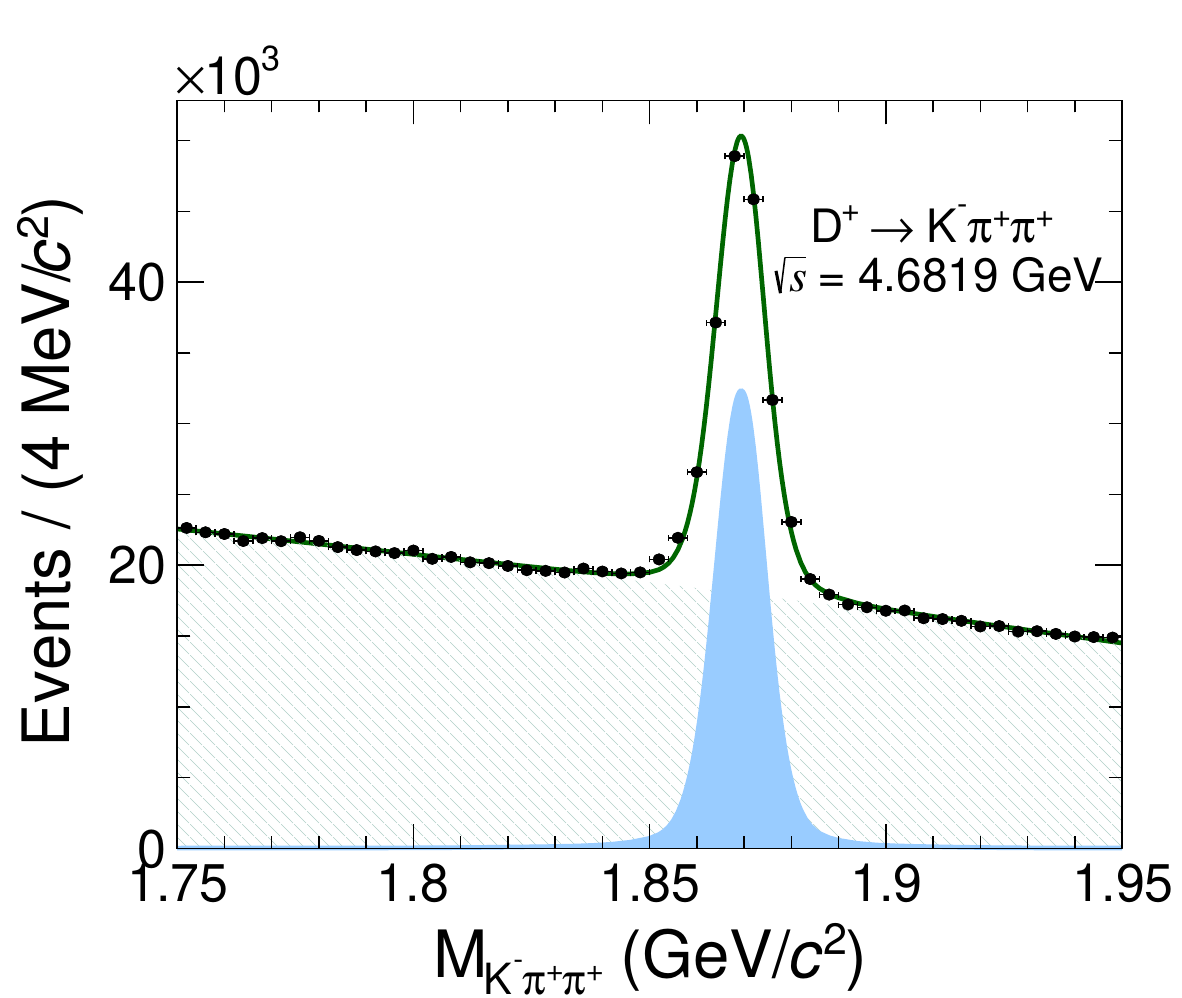}
    \includegraphics[width=0.32\textwidth]{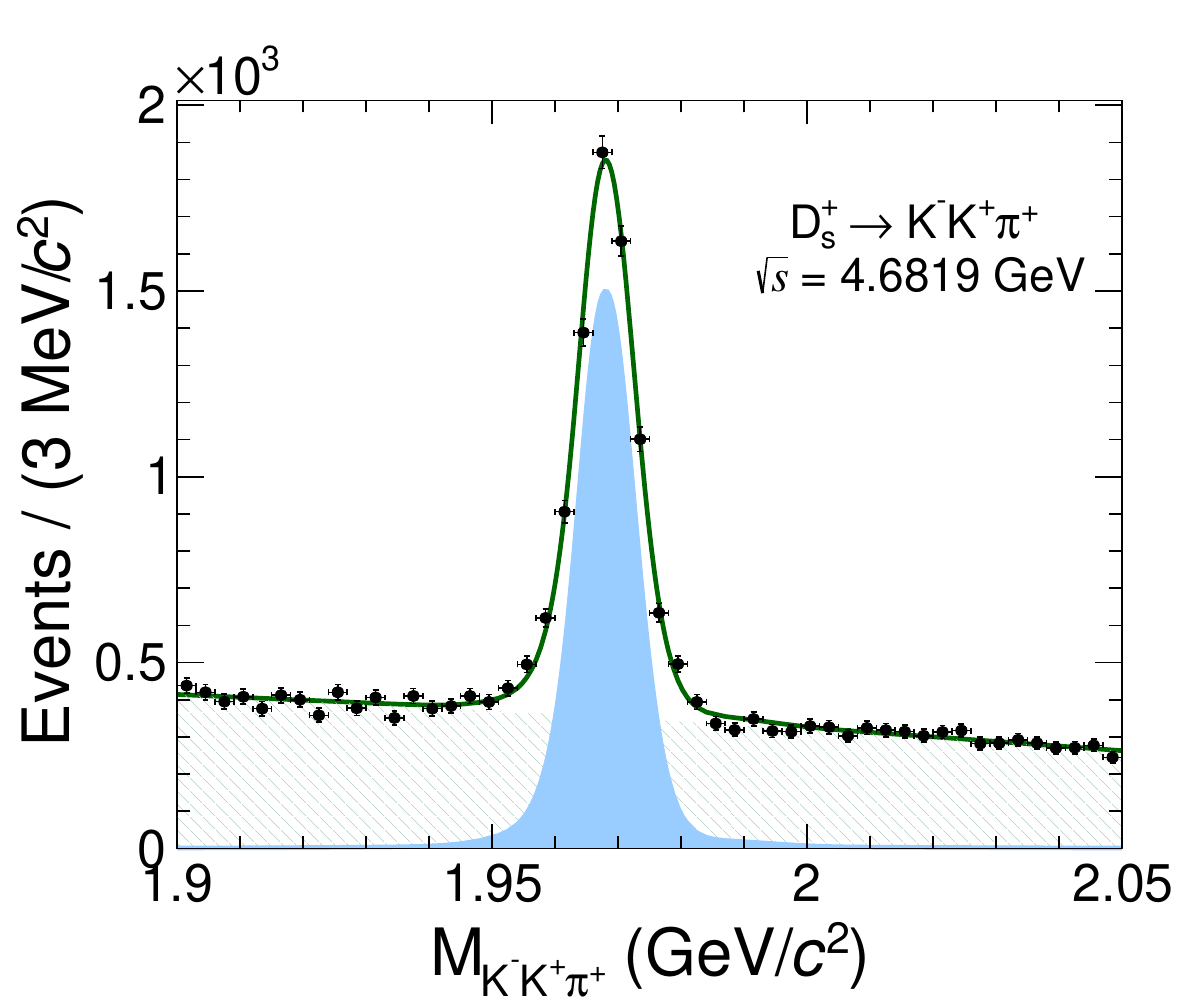}
    \caption{Fit results for $D^{0}\to K^{-}\pi^{+}\pi^{+}\pi^{-}$, $D^{+}\to K^{-}\pi^{+}\pi^{+}$, and $D_{s}^{+}\to K^{-}K^{+}\pi^{+}$ at 4.6819~GeV. The signal and background components are shown as the shaded blue and hatched light-teal regions, respectively, while the total fit is shown as the green curves.}
    \label{eff_for_D}
\end{figure*}

Charged tracks are reconstructed from multilayer drift chamber hits within the   range $\vert\!\cos\theta\vert < 0.93$, where $\theta$ is the track polar angle. Each track is required to have a distance of closest approach to the interaction point within 10 cm along the beam axis and within $1~\text{cm}$ in the transverse plane. Particle identification (PID) combines ${\rm d}E/{\rm d}x$ and time-of-flight  information to calculate the $\chi^2$ for each particle hypothesis and the corresponding $p$-value, $P_i$ ($i=\pi,K$). Kaon (pion) candidates are required to satisfy $P_K>P_\pi$ ($P_\pi>P_K$). Candidate $D^0$, $D^+$, and $D_s^+$ mesons are reconstructed from the selected decay modes. 
A single-tag method is  employed, in which only one signal decay mode is reconstructed per event while the decay of the accompanying charm meson is not considered. As the processes are inclusive and the kinematic distributions of the $D^0$, $D^+$, and $D_s^+$ mesons are not known a priori, all $K^-\pi^+\pi^+\pi^-$, $K^-\pi^+\pi^+$, and $K^-K^+\pi^+$ candidate combinations are retained. MC topology studies~\cite{Zhou:2020ksj} show that events with multiple candidate combinations contribute negligibly to the signal peak, while only increasing the smooth combinatorial background.

For the decay $D^{0}\to K^{-}\pi^{+}\pi^{+}\pi^{-}$, the background from $D^0 \to K_S^0 K^-\pi^+$ with $K_S^0 \to \pi^+\pi^-$ is vetoed by requiring the invariant mass of any $\pi^+\pi^-$ pair, $M_{\pi^+\pi^-}$, to be outside the range $(0.480, 0.515)~\mathrm{GeV}/c^2$. For the decay $D^{+}\to K^{-}\pi^{+}\pi^{+}$, no additional requirements are imposed. For the decay $D^{+}_{s} \to K^{-}K^{+}\pi^{+}$, to improve the signal-to-background ratio, the invariant mass of the $K^-K^+$ pair, $M_{K^+K^-}$,  is required to be close to $M_{\phi}$\cite{BESIII:2024zdh}, {\it i.e.}, $|M_{K^+K^-}-M_{\phi}| < 0.01~\mathrm{GeV}/c^2$, where $M_{\phi}$ is the known $\phi$ meson mass~\cite{ParticleDataGroup:2026aaa}.

To determine the signal yields, unbinned maximum-likelihood fits are performed on the invariant mass distributions of the reconstructed $D^0$, $D^+$, and $D_s^+$ candidates. The signal shape is derived from \textsc{hybrid} MC samples selected by MC truth matching, while the background is described by a second-order  polynomial. For data, the signal shape is additionally convolved with a Gaussian function to account for resolution differences between data and MC simulation. Figure~\ref{eff_for_D} shows representative fit results for the $D^0$, $D^+$, and $D_s^+$ signals in data at an energy point with relatively large integrated luminosity.

The same fitting procedure is applied to the \textsc{hybrid} MC samples to extract the selected signal yields. The yields of generated signals are obtained using the TopoAna tool~\cite{Zhou:2020ksj}. The signal efficiency for each channel is then determined as the ratio of the selected signal yield to the corresponding generated signal yield.

The Born cross section for $e^+e^- \to DX$ (where $D$ denotes $D^0$, $D^+$, or $D_s^+$) is calculated as
\begin{equation}
\begin{aligned}
\sigma_{0} = & \frac{\sigma^{\rm obs}}
{(1+\delta)\cdot |1-\Pi(s)|^{-2}}\\
= & \frac{N^{\rm obs}}
{\mathcal{L}\cdot\epsilon\cdot\mathcal{B}\cdot(1+\delta)\cdot|1-\Pi(s)|^{-2}},
\label{cross_section}
\end{aligned}
\end{equation}
where $\sigma_{0}$ and $\sigma^{\rm obs}$ denote the Born and observed cross sections, respectively. The factor $(1+\delta)$ is the ISR correction factor, and $|1-\Pi(s)|^{-2}$ is the vacuum polarization (VP) factor~\cite{WorkingGrouponRadiativeCorrections:2010bjp}. Here, $N^{\rm obs}$ is the observed number of signal events obtained from fits to the invariant mass spectra, $\mathcal{L}$ is the integrated luminosity~\cite{BESIII:2022ulv,BESIII:2022dxl,BESIII:2017lkp}, $\epsilon$ is the event selection efficiency, and $\mathcal{B}$ is the branching fraction of the tagged decay channel taken from the Particle Data Group~\cite{ParticleDataGroup:2026aaa}. The dressed cross section is defined as $\sigma^{\rm dressed} = \sigma_{0}\cdot |1-\Pi(s)|^{-2}$, where the ISR effects are removed while the VP effects are retained.

\begin{figure*}[htp]
    \centering
     \includegraphics[width=0.7\textwidth]{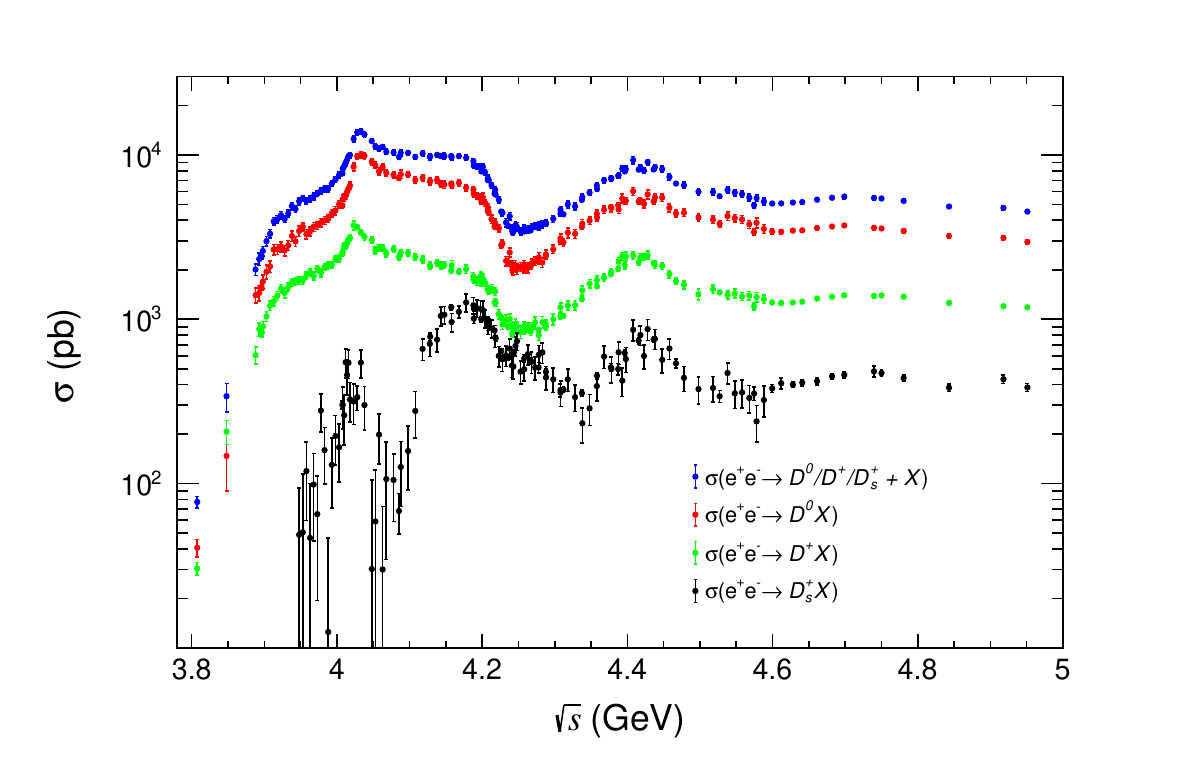} 
     \caption{Comparison of the measured Born cross sections for $\sigma(e^{+}e^{-}\to D^{0}X)$, $\sigma(e^{+}e^{-}\to D^{+}X)$, $\sigma(e^{+}e^{-}\to D^{+}_{s}X)$, and $\sigma(e^{+}e^{-}\to D^{0}/D^{+}/D^{+}_{s} + X)$.}
    \label{compare_born_cross_section}
\end{figure*}

The ISR  factors are determined through an iterative procedure based on the \textsc{hybrid} generator simulation.  The initial ISR factor $(1+\delta)$ is  calculated as
\begin{equation}
1 + \delta =
    \frac{\int W(s,x)\,\sigma_{0}(s(1-x))\,{\rm d}x}
{\sigma_{0}(s)},
\label{isr_factor}
\end{equation}
where $W(s,x)$ is the ISR radiator function~\cite{Nicrosini:1986sm,Benayoun:1999hm,Ping:2016pms,BESIII:2023gqy}, $x = 2E_{\gamma}/\sqrt{s}$ denotes the fraction of energy carried away by ISR photon(s), and $\sigma_{0}$ is the input Born cross section.  
To construct the input cross section line shapes for the iteration, the observed cross sections for $e^+e^- \to D^0X$, $D^+X$, and $D_s^+X$ are obtained by averaging the independent measurements of the tagged decay modes and their charge-conjugate modes. 
To evaluate the ISR factor in Eq.~(\ref{isr_factor}), the resulting observed cross section of $\sigma(e^+e^- \to D^0/D^+/D_s^+ + X)$ is first parameterized and smoothed using variable span smoothing methods~\cite{BESIII:2024zdh,Cleveland1979RobustLW,Cleveland1981RobustLW,Friedman1984SMARTUG,Friedman1984AVS} and taken as the initial input Born cross section. After obtaining the ISR factor together with the VP factor, the resulting Born cross section is reparameterized and smoothed to serve as the input line shape for subsequent iterations in the \textsc{hybrid} simulation. In each iteration, the production fractions of known and unknown open-charm processes are updated accordingly. Using the updated line shape and corresponding production fractions, the \textsc{hybrid} simulation is re-run at 150 c.m. energy points, and the selection efficiencies are recalculated based on the new simulation results. The updated Born cross sections are then used as input for the next iteration.  This procedure is repeated until the Born cross sections converge. 
For the individual channels, \(e^+e^-\to D^0X\), \(D^+X\), and \(D_s^+X\), the ISR  factors are calculated separately using Eq.~(\ref{isr_factor}). For the total cross section, a single ISR  factor obtained from the \textsc{hybrid} MC simulation is used. The sum of the three individually corrected Born cross sections is consistent with the total Born cross section obtained using this single ISR  factor, with negligible differences. 
Figure~\ref{compare_born_cross_section} shows the measured Born cross sections for $e^+e^- \to D^0 X$, $e^+e^- \to D^+ X$, $e^+e^- \to D_s^+ X$, and $e^{+}e^{-}\to D^{0}/D^{+}/D^{+}_{s} + X$.

The systematic uncertainties of the measured cross sections arise from various sources. 
The uncertainties associated with tracking and PID are evaluated by comparing the $K/\pi$ efficiencies between data and MC using control samples of the process $e^{+}e^{-} \to K^{+}K^{-}\pi^{+}\pi^{-}$. For each final-state track from the $D^0$, $D^+$, and $D_s^+$ decays, the corresponding uncertainty is determined by momentum-dependent reweighting according to the data--MC efficiency difference in each momentum interval. The total uncertainty is then obtained by  linearly summing the contributions from all final-state tracks. 
The uncertainties associated with the $\phi$ mass requirement, the $K_S^0$ veto, and the fitting procedure are evaluated by repeating the analysis with several reasonable alternative selections. For the $\phi$ mass requirement, the mass window is varied from $[1.01,1.03]~\mathrm{GeV}/c^2$ to $[1.005,1.035]$, $[1.00,1.04]$, and $[0.995,1.045]~\mathrm{GeV}/c^2$. For the $K_S^0$ veto, the excluded mass region is varied to $0.475 < M_{\pi^+\pi^-} < 0.520~\mathrm{GeV}/c^2$ and $0.485 < M_{\pi^+\pi^-} < 0.510~\mathrm{GeV}/c^2$. To assess the stability of the signal yield extraction under variations in the fitting procedure, the fit range is varied within reasonable limits. The nominal ranges for $D^0$ and $D^+$ are varied from $[1.75,1.95]~\mathrm{GeV}/c^2$ to $[1.73,1.97]$, $[1.74,1.96]$, $[1.76,1.94]$, $[1.77,1.93]$, $[1.78,1.92]$, and $[1.79,1.91]~\mathrm{GeV}/c^2$, while the nominal range for $D_s^+$ is varied from $[1.90,2.05]~\mathrm{GeV}/c^2$ to $[1.88,2.07]$, $[1.89,2.06]$, $[1.91,2.04]$, $[1.92,2.03]$, $[1.93,2.02]$, and $[1.94,2.01]~\mathrm{GeV}/c^2$. The largest deviations from the nominal results are assigned as the corresponding systematic uncertainties. 
The uncertainty in the integrated luminosity is determined using Bhabha events and is estimated to be 1.00\%~\cite{BESIII:2022ulv,BESIII:2022dxl,BESIII:2017lkp}.  The uncertainties in the branching fractions of $D^{0}\to K^{-}\pi^{+}\pi^{+}\pi^{-}$, $D^{+}\to K^{-}\pi^{+}\pi^{+}$, and $D^{+}_{s}\to K^{-}K^{+}\pi^{+}$ are 1.70\%, 1.71\%, and 1.47\% \cite{ParticleDataGroup:2026aaa}, respectively. The uncertainty associated with the background description is evaluated by replacing the nominal second-order  polynomial with a third-order polynomial.

The uncertainty associated with the ISR correction is evaluated from two sources: one arising from the intrinsic precision of the calculation method and the other originating from the uncertainty in the input Born cross sections. To estimate the intrinsic precision of the calculation method, an alternative approach is adopted in which the ISR factor in Eq.~(\ref{isr_factor}) is calculated directly, rather than being obtained from the \textsc{hybrid} simulation. The Born cross sections derived using these two independent methods are compared, and the difference between them is taken as the first contribution to the ISR uncertainty. To evaluate the uncertainty arising from the input Born cross sections, the ISR factors obtained using parameterized and interpolated Born cross section line shapes are compared. The difference between the resulting Born cross sections is taken as the second contribution to the ISR uncertainty. The total uncertainty of the ISR correction is then obtained by summing the two contributions in quadrature.   
For each individual channel, the ISR factor is calculated directly using Eq.~(\ref{isr_factor}). The corresponding ISR uncertainty is obtained by combining in quadrature the uncertainty associated with the parameterized and interpolated line shapes and the method uncertainty determined from the total cross section analysis described above.

The total systematic uncertainty is obtained by adding all individual contributions in quadrature. The total uncertainty is determined by combining the statistical and total systematic uncertainties in quadrature. 
The final cross sections for $e^+e^- \to D^0X$, $e^+e^- \to D^+X$, and $e^+e^- \to D_s^+X$ are obtained by combining the measurements of the tagged decay modes and their charge-conjugate modes~\cite{Schmelling:1994pz,BESIII:2017kqg}.
The uncertainties of these combined cross sections account for correlations between the systematic uncertainties of each tagged decay mode and its charge-conjugate mode~\cite{BESIII:2017kqg}.
The total cross section for  $e^+e^- \to D^{0}/D^{+}/D_s^{+} + X$  and its uncertainty are then obtained by combining the three individual cross sections, with correlations among the systematic uncertainties considered. 
Table~\ref{sys_all_sum_D0_Dp_Ds} summarizes the relative systematic uncertainties for $\sigma(e^+e^- \to D^0X)$, $\sigma(e^+e^- \to D^+X)$, $\sigma(e^+e^- \to D^+_sX)$, and  $\sigma(e^+e^- \to D^0/ D^+/D^+_s + X)$. These values represent the typical magnitudes of the uncertainties over the full energy range and are provided for reference, as the systematic uncertainties vary across different energy points. Table~\ref{all_sum_D0_Dp_Ds_value} presents the total cross section $\sigma(e^+e^- \to D^0/ D^+/D^+_s + X)$ at the high c.m. energies.  Detailed values at each energy point, together with the full covariance matrix, are provided in the Supplemental Material~\cite{supp_ma}.

Figure~\ref{compare_born_cross_section} shows the measured Born cross sections for the different channels with total uncertainties.  The cross section for $e^+e^- \to D^0 X$ is the largest, followed by $e^+e^- \to D^+ X$, while $e^+e^- \to D_s^+ X$ is the smallest. The $D^0 X$ and $D^+ X$ channels exhibit similar line shapes, whereas the $D_s^+ X$ channel shows a different behavior.  These line shapes reveal possible enhancements corresponding to the $Y(4500)$~\cite{Wang:2025zbv} and $Y(4700)$~\cite{BESIII:2023wqy}.

\begin{table}[htp]
    \caption{Summary of the  relative systematic uncertainties for $\sigma(e^+e^- \to D^0X)$, $\sigma(e^+e^- \to D^+X)$, $\sigma(e^+e^- \to D^+_sX)$, and  $\sigma(e^+e^- \to D^0/ D^+/D^+_s + X)$. Detailed values at each energy point are given in the Supplemental Material~\cite{supp_ma}.}
     \renewcommand*{\arraystretch}{1.5}
    \setlength{\tabcolsep}{0.5mm}{
    \begin{tabular}{l|c c c c}
    \hline
 Source  &   $ \delta\sigma_{D^{0}}$(\%) & $\delta\sigma_{D^{+}}$(\%)  &$\delta\sigma_{D^{+}_s}$(\%) & $\delta\sigma_{\rm tot}$(\%)\\
    \hline  

    Tracking                      & 0.72           & 0.59  & 1.09    & 0.72  \\                                   
      PID                         & 1.54           & 0.90  & 1.87    & 1.41  \\                                        
      $\phi$ mass requirement     & --             & --    & 1.05    & 0.12 \\                                        
      $K_S^0$ veto                & 0.19           & --    & --    & 0.13 \\                                 
      $\mathcal{L}$               & 1.00           & 1.00  & 1.00    & 1.00 \\                                        
      $\mathcal{B}$               & 1.70           & 1.71  & 1.47    & 1.21 \\                                       
      ISR factor                  & 2.17           & 2.25  & 2.16    & 0.61 \\                     
      Fitting procedure           & 0.63           & 0.36  & 1.14    & 0.46 \\               
      Background shape            & 0.17           & 0.32  & 0.39    & 0.15  \\  
          \hline
      Total                       & 3.46            & 3.22   & 3.88     & 2.37  \\

    \hline
    \end{tabular}}
    \label{sys_all_sum_D0_Dp_Ds}
\end{table}

\begin{table*}[htp]
\caption{ Summary of the $\sigma(e^+e^- \to D^0/D^+/D^+_s + X )$ at the high energy points.  The columns correspond to the c.m. energy $\sqrt{s}$, the observed cross section (Obs.), its statistical uncertainty (Stat.\ unc.), the ISR correction factor $(1+\delta)$, the dressed cross section, the VP correction factor $|1-\Pi(s)|^{-2}$, the Born cross section with its statistical and systematic uncertainties, and the relative systematic uncertainty (Rel.\ Sys.).  Detailed values at each energy point are given in the Supplemental Material~\cite{supp_ma}. } 	
  \renewcommand*{\arraystretch}{1.5}
    \setlength{\tabcolsep}{0.5mm}{
    \begin{tabular}{l|c c c c c c c c c c}
\hline
\centering
 $\sqrt{s}$ (GeV) & Obs. ($\rm pb$) & Stat. unc. ($\rm pb$) & $(1+\delta)$ & Dressed ($\rm pb$) & $|1-\Pi(s)|^{-2}$ & Born $\pm$ Stat. $\pm$ Sys. ($\rm pb$) & Rel. Sys. (\%)\\
\hline

4.6119           & 5334           & 56             & 0.9984         & 5342           & 1.0545         & 5066            $\pm$    53 $\pm$ 119             & 2.36       \\  
4.6280           & 5368           & 25             & 0.9934         & 5404           & 1.0544         & 5125            $\pm$    24 $\pm$ 115             & 2.25       \\  
4.6409           & 5382           & 24             & 0.9892         & 5441           & 1.0544         & 5160            $\pm$    23 $\pm$ 118             & 2.28       \\  
4.6612           & 5533           & 25             & 0.9832         & 5628           & 1.0544         & 5337            $\pm$    24 $\pm$ 125             & 2.33       \\  
4.6819           & 5661           & 15             & 0.9797         & 5778           & 1.0545         & 5480            $\pm$    14 $\pm$ 130             & 2.37       \\  
4.6988           & 5746           & 26             & 0.9777         & 5877           & 1.0545         & 5573            $\pm$    25 $\pm$ 173             & 3.11       \\  
4.7397           & 5651           & 46             & 0.9796         & 5768           & 1.0548         & 5468            $\pm$    45 $\pm$ 133             & 2.43       \\  
4.7500           & 5616           & 31             & 0.9811         & 5725           & 1.0549         & 5427            $\pm$    30 $\pm$ 126             & 2.33       \\  
4.7805           & 5451           & 26             & 0.9849         & 5534           & 1.0552         & 5245            $\pm$    25 $\pm$ 122             & 2.32       \\  
4.8431           & 5093           & 25             & 0.9954         & 5117           & 1.0557         & 4847            $\pm$    24 $\pm$ 111             & 2.30       \\  
4.9180           & 5054           & 42             & 1.0064         & 5022           & 1.0562         & 4755            $\pm$    39 $\pm$ 114             & 2.39       \\  
4.9509           & 4824           & 46             & 1.0112         & 4771           & 1.0564         & 4516            $\pm$    43 $\pm$ 102             & 2.26       \\  

    \hline
    \end{tabular}}
    \label{all_sum_D0_Dp_Ds_value}
\end{table*}

\begin{figure*}[htp]
\centering
\begin{overpic}[width=0.70\textwidth]{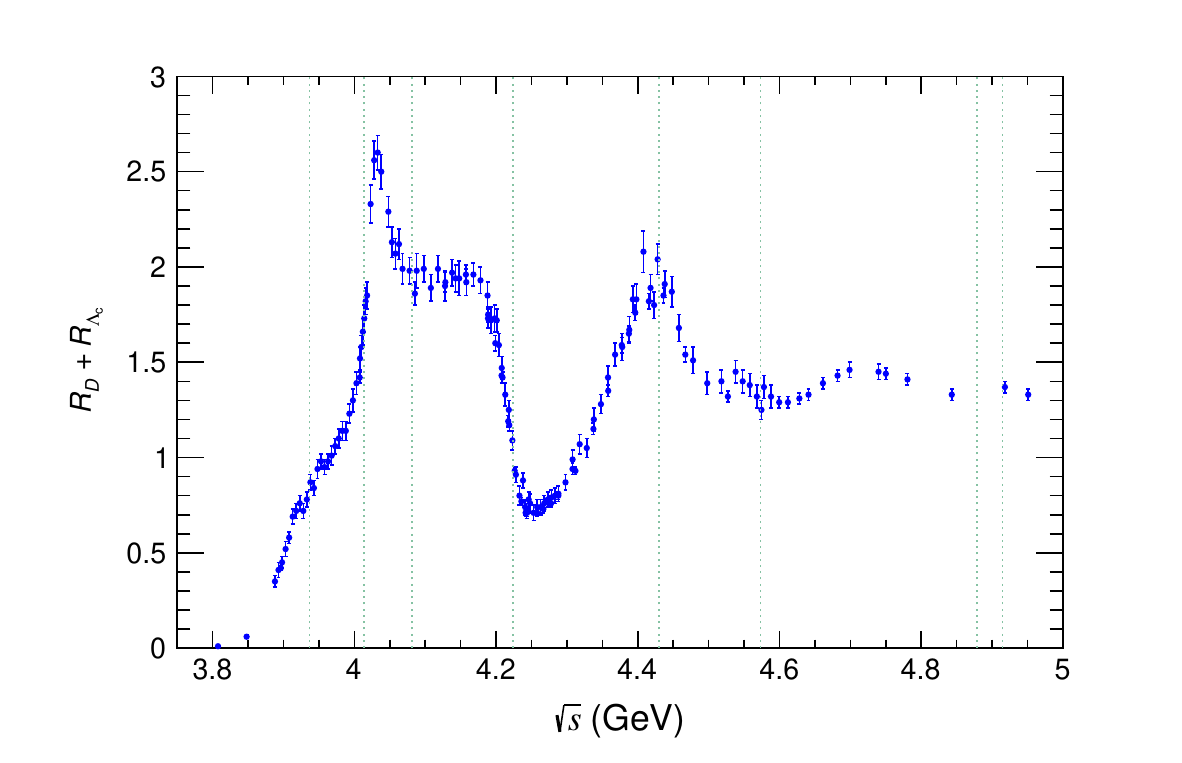}
\put(25,13){\rotatebox{90}{\scriptsize $D_s^{+}D_s^{-}$}}
\put(30,13){\rotatebox{90}{\scriptsize $D^{*0}\bar{D}^{*0}$}}
\put(33,13){\rotatebox{90}{\scriptsize $D_s^{+}D_s^{*-}$}}
\put(42,13){\rotatebox{90}{\scriptsize $D_s^{*+}D_s^{*-}$}}
\put(54,13){\rotatebox{90}{\scriptsize $D_s^{*+}D_{s0}^{*}(2317)^{-}$}}
\put(63,13){\rotatebox{90}{\scriptsize $\Lambda_c^{+}\bar{\Lambda}_c^{-}$}}
\put(81,13){\rotatebox{90}{\scriptsize $\Lambda_c^{+}\bar{\Lambda}_c(2595)^{-}$}}
\put(85,13){\rotatebox{90}{\scriptsize $\Lambda_c^{+}\bar{\Lambda}_c(2625)^{-}$}}
\end{overpic}
\caption{Distribution of the $R_D+R_{\Lambda_c}$ values. The green dashed lines indicates the two-body thresholds.}
\label{compare_Rc}
\end{figure*}

To characterize the contribution of inclusive open-charm production to the total hadronic cross section, the ratio
\begin{equation}
R_{D} = \frac{\sigma(e^{+}e^{-}\to D^{0}/D^{+}/D_{s}^{+} + X)}{\sigma(e^{+}e^{-}\to \mu^{+}\mu^{-})}
\end{equation}
is determined. The analogous ratios $R_{\Lambda_c^+\bar{\Lambda}_c^-}$, $R_{\Lambda_c^+\bar{\Lambda}_c(2595)^-}$, and $R_{\Lambda_c^+\bar{\Lambda}_c(2625)^-}$ for the corresponding charmed-baryon production processes~\cite{BESIII:2017kqg,BESIII:2023rwv,BESIII:2023eie} are also determined above their respective thresholds near $4.5745$ and $4.9180~\text{GeV}$, as these channels contribute significantly to the open-charm yield at higher energies. The summed charmed-baryon contribution is defined as $R_{\Lambda_c} = R_{\Lambda_c^+\bar{\Lambda}_c^-} + R_{\Lambda_c^+\bar{\Lambda}_c(2595)^-} + R_{\Lambda_c^+\bar{\Lambda}_c(2625)^-}$. As shown in Fig.~\ref{compare_Rc}, distinct enhancements are observed near $4.5$ and $4.7~\text{GeV}$, corresponding to the $Y(4500)$ and $Y(4700)$ structures, respectively. In comparison, the contributions from known hidden-charm processes are substantially smaller than those from open-charm meson production in this energy region. This confirms that inclusive open-charm meson production constitutes the dominant hadronic contribution above threshold, providing a reliable approximation to the total hadronic cross section.

If the light-quark contribution $R_{uds}^{\rm pQCD}\simeq2.139$~\cite{Boito:2025qfz} is added to the measured sum $R_D+R_{\Lambda_c}$, the resulting total $R$ value is comparable to, and in some energy regions even slightly exceeds, the inclusive hadronic $R$ measured by BESII~\cite{BES:2001ckj}, while achieving significantly improved precision. This observation indicates that only a limited fraction of the hadronic cross section remains unmeasured for hidden-charm production above the open-charm threshold. Together with the measured charmed-baryon contributions, the present results therefore place stringent constraints on the overall contribution of hidden-charm and other unmeasured final states~\cite{Farrar:2023zmj}, particularly given that many exotic charmonium-like $Y$ states are observed primarily through hidden-charm final states.  Moreover, since the charm-threshold region contributes to dispersion-relation evaluations of the hadronic vacuum polarization, the present measurement of $R_D$ provides valuable experimental input for  precision determinations of the anomalous magnetic moment of the muon $a_\mu$ and the running electromagnetic coupling constant $\alpha$~\cite{Zhao:2000ev,Huang:2022zqh}.

The results presented here also provide essential input for $K$-matrix-based coupled-channel analyses of the vector charmonium spectrum. Such analyses benefit from incorporating an  inclusive  $e^+e^-\to c\bar c$  channel into the fit, both to improve the quality of the description and to impose additional constraints on the resonance parameters~\cite{Husken:2024hmi}; previously, this inclusive contribution was obtained indirectly by subtracting the light-quark contribution from the total $R$ value. The present measurement of $R_D$, together with $R_{\Lambda_c}$, thus provides a direct and more precise determination of the inclusive open-charm contribution, playing an analogous role to the inclusive $R_{b\bar{b}}$ measurement in the bottomonium analysis of Ref.~\cite{Husken:2022yik}. Furthermore, comparisons between $R_D$ and the sum of exclusive channels can constrain contributions from potentially missing open-charm final states, which are typically accounted for by a dummy channel in coupled-channel fits~\cite{Husken:2022yik,Husken:2024hmi}. Such constraints are expected to reduce ambiguities arising from incomplete channel information, thereby improving the reliability of future determinations of resonance pole positions and channel couplings of vector charmonium and charmonium-like states.

In summary, precise measurements of the inclusive cross sections of open-charm mesons $\sigma(e^{+}e^{-}\to D^{0}X)$, $\sigma(e^{+}e^{-}\to D^{+}X)$, $\sigma(e^{+}e^{-}\to D_{s}^{+}X)$, and their combination $\sigma(e^{+}e^{-}\to D^{0}/D^{+}/D_{s}^{+}+X)$ are presented at 150 c.m. energy points between $3.80$ and $4.95~\mathrm{GeV}$ for the first time. The measured line shapes exhibit clear structures, including enhancements near $4.5$ and $4.7~\mathrm{GeV}$ that provide  evidence for the coupling of the $Y(4500)$ and $Y(4700)$ states to open-charm final states. The enhancement near $4.5~\mathrm{GeV}$ remains visible even after accounting for the $e^{+}e^{-}\to\Lambda_c^{+}\bar{\Lambda}_c^{-}$ threshold  contribution~\cite{BESIII:2017kqg,BESIII:2023rwv}. As one of the most precise measurements of inclusive open-charm production, these results establish a benchmark for the dominant hadronic contribution above the open-charm threshold and provide essential input for understanding open-charm production dynamics and for future coupled-channel studies of the vector charmonium and charmonium-like spectrum.

\acknowledgments
\input{acknowledgement_2026-07-10}


\normalem  
\bibliography{ref}

\onecolumngrid
\vspace*{1em}
\begin{center}
\rule{0.2\textwidth}{1.5pt}
\end{center}
\vspace{1em}

{
\centering
\input{authorlist_2026-07-10.tex}
}

\clearpage
\input{Supplemental_Material.tex}

\end{document}

%% file: acknowledgement_2026-07-10.tex
\textbf{Acknowledgement}

The BESIII Collaboration thanks the staff of BEPCII (https://cstr.cn/31109.02.BEPC) and the IHEP computing center for their strong support. This work is supported in part by National Key R\&D Program of China under Contracts Nos. 2025YFA1613900, 2023YFA1606000, 2023YFA1606704; National Natural Science Foundation of China (NSFC) under Contracts Nos. 11635010, 11935015, 11935016, 11935018, 12025502, 12035009, 12035013, 12061131003, 12192260, 12192261, 12192262, 12192263, 12192264, 12192265, 12221005, 12225509, 12235017, 12342502, 12361141819, 12535005; the Chinese Academy of Sciences (CAS) Large-Scale Scientific Facility Program; the Strategic Priority Research Program of Chinese Academy of Sciences under Contract No. XDA0480600; CAS under Contract No. YSBR-101; 100 Talents Program of CAS; The Institute of Nuclear and Particle Physics (INPAC) and Shanghai Key Laboratory for Particle Physics and Cosmology; Agencia Nacional de Investigaci\'on y Desarrollo de Chile (ANID), Chile under Contract No. ANID CCTVal CIA250027; Istituto Nazionale di Fisica Nucleare, Italy; Knut and Alice Wallenberg Foundation under Contracts Nos. 2021.0174, 2021.0299, 2023.0315; Ministry of Development of Turkey under Contract No. DPT2006K-120470; National Research Foundation of Korea under Contract No. RS-2026-25486791; National Science and Technology fund of Mongolia; Polish National Science Centre under Contracts Nos. 2024/53/B/ST2/00975, 2025/58/E/ST2/00051; STFC (United Kingdom); Swedish Research Council under Contract No. 2019.04595; U. S. Department of Energy under Contract No. DE-FG02-05ER41374



%% file: authorlist_2026-07-10.tex
M.~Ablikim$^{1}$\BESIIIorcid{0000-0002-3935-619X},
M.~N.~Achasov$^{4,c}$\BESIIIorcid{0000-0002-9400-8622},
P.~Adlarson$^{85}$\BESIIIorcid{0000-0001-6280-3851},
X.~C.~Ai$^{91}$\BESIIIorcid{0000-0003-3856-2415},
C.~S.~Akondi$^{32A,32B}$\BESIIIorcid{0000-0001-6303-5217},
R.~Aliberti$^{40}$\BESIIIorcid{0000-0003-3500-4012},
A.~Amoroso$^{84A,84C}$\BESIIIorcid{0000-0002-3095-8610},
Q.~An$^{80,67,\dagger}$,
M.~S.~Anderson$^{40}$\BESIIIorcid{0009-0008-1550-2632},
Y.~Bai$^{65}$\BESIIIorcid{0000-0001-6593-5665}
% ...
\\
\vspace{0.2cm}
(BESIII Collaboration)\\
\vspace{0.2cm} {\it
$^{1}$ Institute of High Energy Physics, Beijing 100049, People's Republic of China\\
% ...
$^{4}$ Budker Institute of Nuclear Physics SB RAS (BINP), Novosibirsk 630090, Russia\\
% ...
$^{32}$ INFN Sezione di Ferrara, (A)INFN Sezione di Ferrara, I-44122, Ferrara, Italy; (B)University of Ferrara, I-44122, Ferrara, Italy\\
% ...
$^{40}$ Johannes Gutenberg University of Mainz, Johann-Joachim-Becher-Weg 45, D-55099 Mainz, Germany\\
% ...
$^{65}$ Southeast University, Nanjing 211100, People's Republic of China\\
% ...
$^{67}$ State Key Laboratory of Particle Detection and Electronics, Beijing 100049, Hefei 230026, People's Republic of China\\
% ...
$^{80}$ University of Science and Technology of China, Hefei 230026, People's Republic of China\\
% ...
$^{84}$ University of Turin and INFN, (A)University of Turin, I-10125, Turin, Italy; (B)University of Eastern Piedmont, I-15121, Alessandria, Italy; (C)INFN, I-10125, Turin, Italy\\
$^{85}$ Uppsala University, Box 516, SE-75120 Uppsala, Sweden\\
% ...
$^{91}$ Zhengzhou University, Zhengzhou 450001, People's Republic of China\\

\vspace{0.2cm}
$^{\dagger}$ Deceased\\
% ...
$^{c}$ Also at the Novosibirsk State University, Novosibirsk, 630090, Russia\\
% ...

}

%% file: Supplemental_Material.tex
















\section*{Supplemental Material}

\begin{figure*}[htp]
    \centering
    \includegraphics[width=0.35\textwidth]{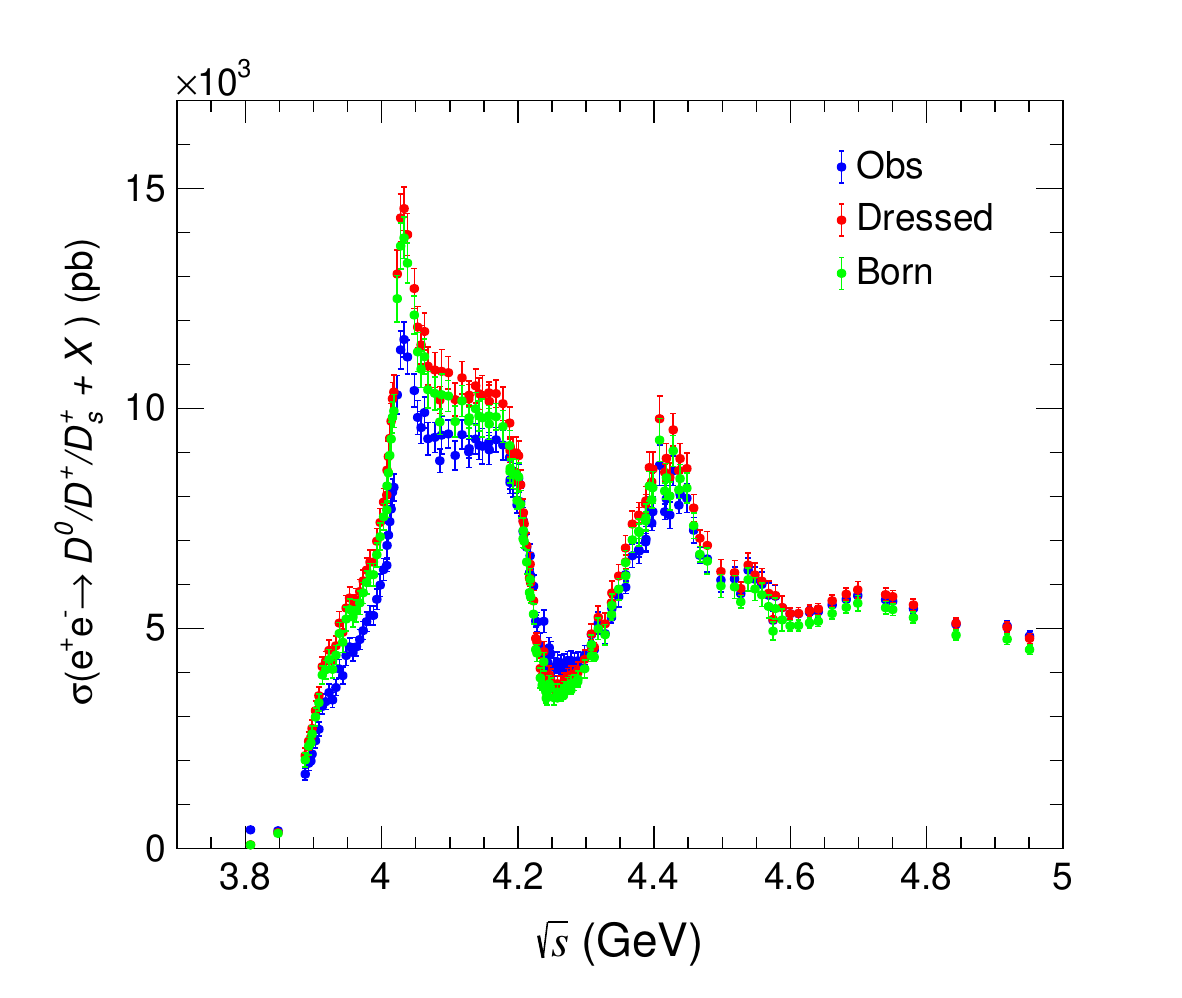}
    \includegraphics[width=0.35\textwidth]{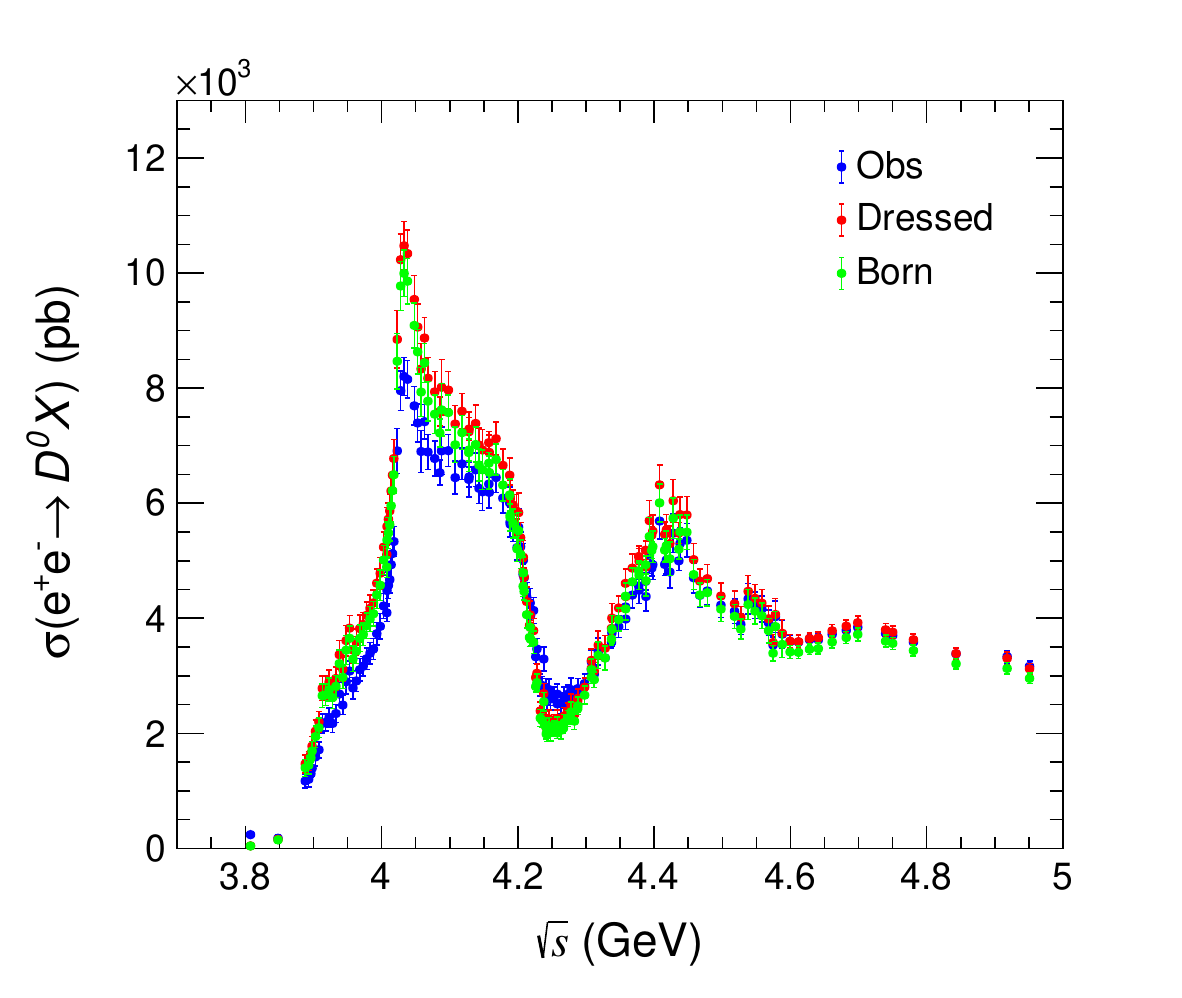}\\
    \includegraphics[width=0.35\textwidth]{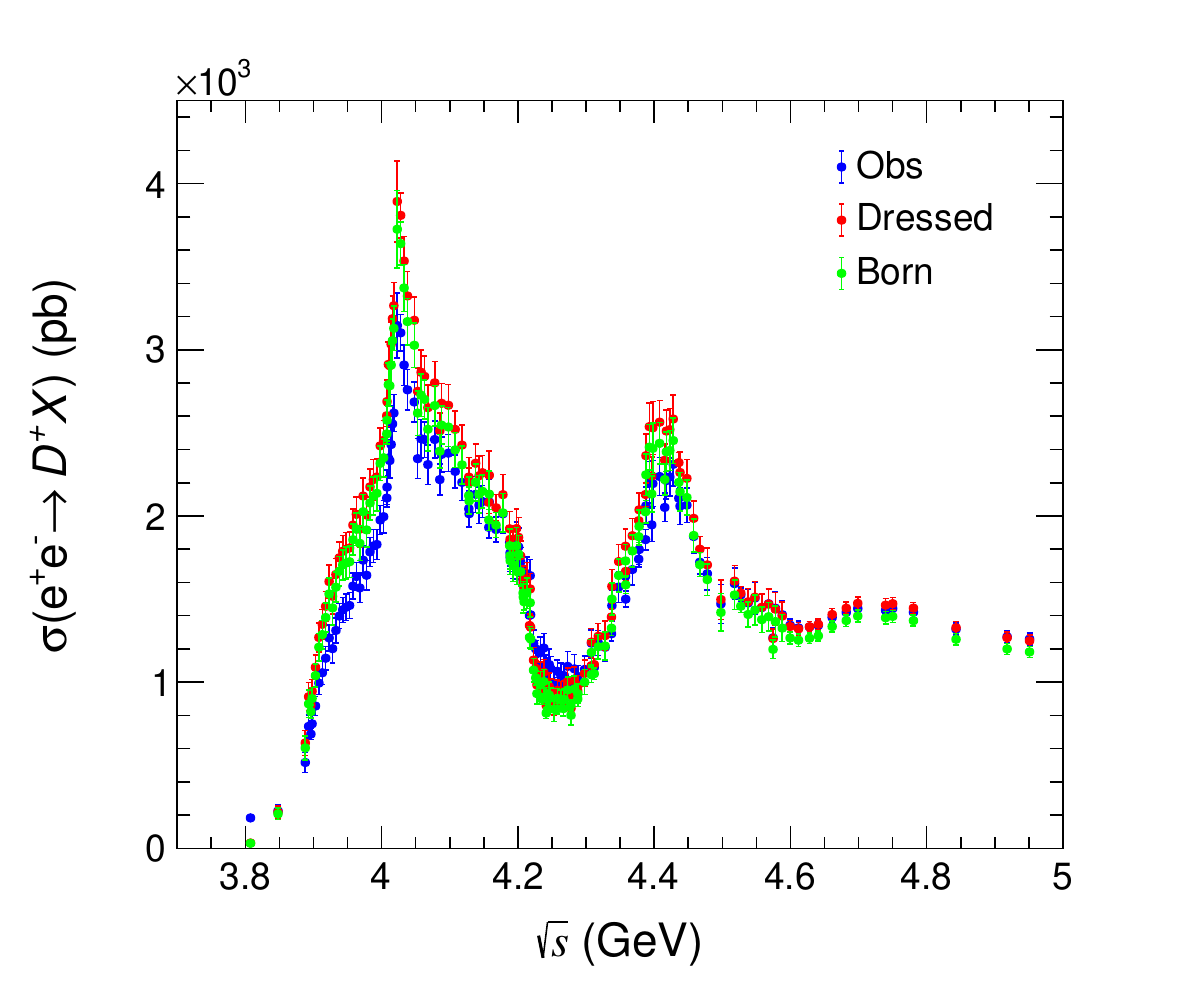}
    \includegraphics[width=0.35\textwidth]{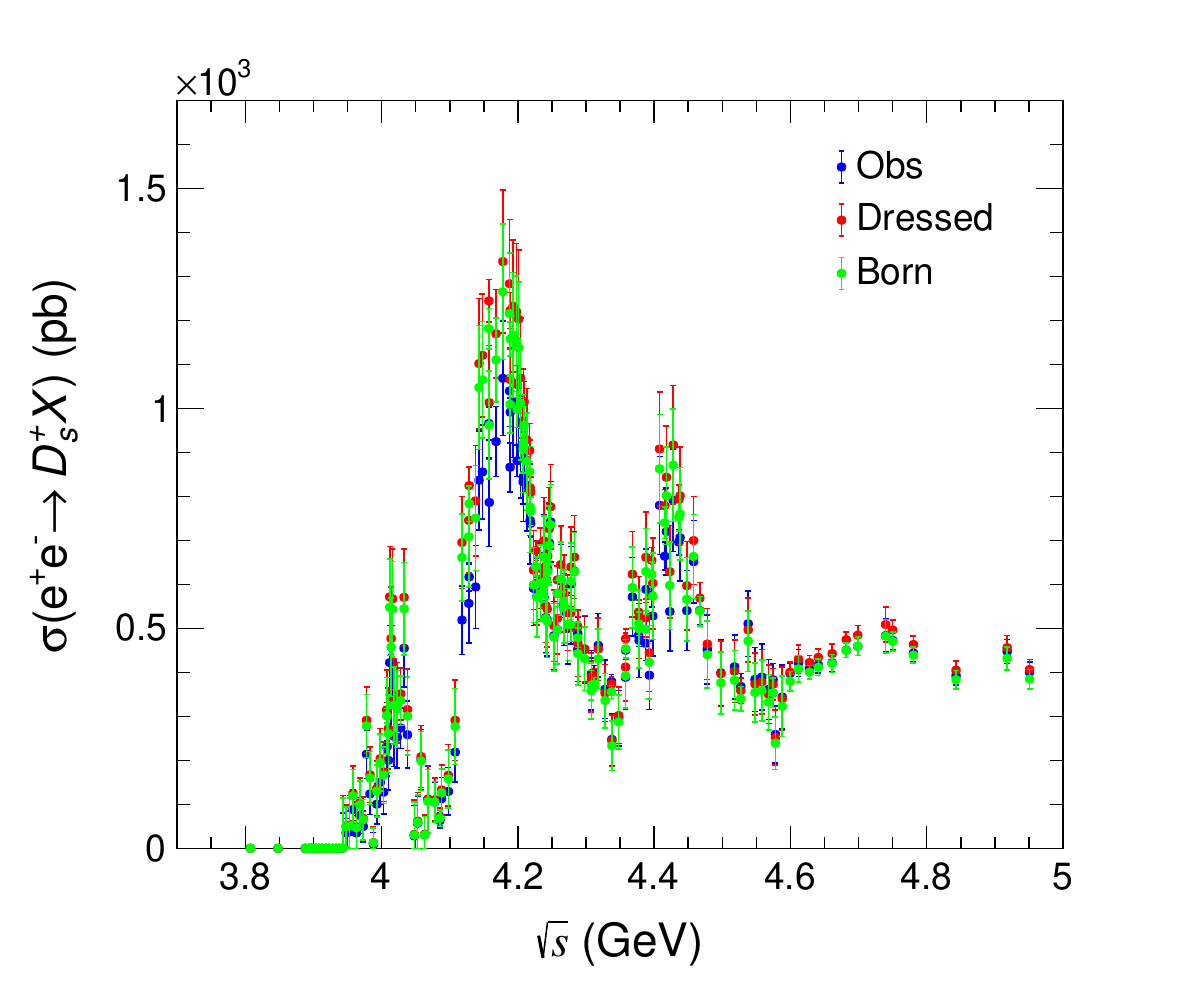}
    
    \caption{The observed, dressed, and Born cross sections for \( \sigma(e^{+}e^{-}\to D^{0}/D^{+}/D_{s}^{+} + X) \), \( \sigma(e^{+}e^{-}\to D^{0}X) \), \( \sigma(e^{+}e^{-}\to D^{+}X) \), and \( \sigma(e^{+}e^{-}\to D_{s}^{+}X) \) are presented with their total uncertainties. The corresponding values are listed in Tables~\ref{allD_obs_isr_fvp}, \ref{D0_channel_cross_section}, \ref{Dp_channel_cross_section}, and \ref{Ds_channel_cross_section}, respectively.}
    \label{sys_D0_Dp_Dsp_obs_dress_born}
\end{figure*}

Figure~\ref{sys_D0_Dp_Dsp_obs_dress_born} shows the observed, dressed, and Born cross sections for \( \sigma(e^{+}e^{-}\to D^{0}/D^{+}/D_{s}^{+} + X) \), \( \sigma(e^{+}e^{-}\to D^{0}X) \), \( \sigma(e^{+}e^{-}\to D^{+}X) \), and \( \sigma(e^{+}e^{-}\to D_{s}^{+}X) \), together with their total uncertainties.

Tables~\ref{allD_obs_isr_fvp}, \ref{D0_channel_cross_section}, \ref{Dp_channel_cross_section}, and \ref{Ds_channel_cross_section} summarize the measured cross sections for $e^{+}e^{-}\to D^{0}/D^{+}/D_{s}^{+}+X$, $e^{+}e^{-}\to D^{0}X$, $e^{+}e^{-}\to D^{+}X$, and $e^{+}e^{-}\to D_{s}^{+}X$, respectively. The columns correspond to the center-of-mass energy $\sqrt{s}$, the observed cross section (Obs.), its statistical uncertainty (Stat.\ unc.), the ISR correction factor $(1+\delta)$, the dressed cross section, the VP correction factor $|1-\Pi(s)|^{-2}$, the Born cross section with its statistical and systematic uncertainties, and the relative systematic uncertainty (Rel.\ Sys.).

Table~\ref{summary of RD} summarizes the measured $R_D$ values. Table~\ref{summary_energy_lumi_xyz_rscan} summarizes the integrated luminosities of all data samples.

\clearpage

\begin{longtable}[h]{c  c  c  c c  c c c c}
\caption{ Summary of $\sigma(e^+e^- \to D^0/D^+/D^+_s + X )$.  (150 points)} 	\\
\hline
\centering
 $\sqrt{s}$ (GeV) & Obs. ($\rm pb$) & Stat. unc. ($\rm pb$) & $(1+\delta)$ & Dressed ($\rm pb$) & $|1-\Pi(s)|^{-2}$ & Born $\pm$ Stat. $\pm$ Sys. ($\rm pb$) & Rel. Sys. (\%)\\
\hline
\endfirsthead  
\hline
  $\sqrt{s}$ (GeV) & Obs. ($\rm pb$) & Stat. unc. ($\rm pb$) & $(1+\delta)$ & Dressed ($\rm pb$) & $|1-\Pi(s)|^{-2}$ & Born $\pm$ Stat. $\pm$ Sys. ($\rm pb$) & Rel. Sys. (\%)\\
\hline
\endhead 
\hline
\endfoot 
\endlastfoot 


3.8077           & 421            & 30             & 5.1575         & 82             & 1.0562         & 77              $\pm$     5 $\pm$ 3               & 4.29       \\  
3.8480           & 400            & 76             & 1.1168         & 358            & 1.0521         & 340             $\pm$    65 $\pm$ 16              & 4.83       \\  
3.8880           & 1687           & 125            & 0.8004         & 2108           & 1.0494         & 2009            $\pm$   148 $\pm$ 67              & 3.33       \\  
3.8930           & 1934           & 133            & 0.7937         & 2437           & 1.0489         & 2323            $\pm$   160 $\pm$ 123             & 5.28       \\  
3.8962           & 1987           & 46             & 0.7889         & 2518           & 1.0487         & 2401            $\pm$    56 $\pm$ 57              & 2.37       \\  
3.8980           & 2143           & 123            & 0.7862         & 2726           & 1.0485         & 2599            $\pm$   150 $\pm$ 95              & 3.66       \\  
3.9030           & 2446           & 133            & 0.7817         & 3129           & 1.0483         & 2985            $\pm$   162 $\pm$ 125             & 4.20       \\  
3.9080           & 2707           & 136            & 0.7803         & 3470           & 1.0481         & 3310            $\pm$   166 $\pm$ 100             & 3.02       \\
\end{longtable}

\clearpage
\begin{longtable}[h]{c  c  c  c c  c c c c}
\caption{Summary of $\sigma(e^+e^- \to D^0X)$. The ISR factor values are calculated directly from the formula rather than from the \textsc{hybrid} generator simulation. (150 points)} 	\\
\hline
\centering
$\sqrt{s}$ (GeV) & Obs. ($\rm pb$) & Stat. unc. ($\rm pb$) & $(1+\delta)$ & Dressed ($\rm pb$) & $|1-\Pi(s)|^{-2}$ & Born $\pm$ Stat. $\pm$ Sys. ($\rm pb$) & Rel. Sys. (\%)\\
\hline
\endfirsthead 
\hline          
$\sqrt{s}$ (GeV) & Obs. ($\rm pb$) & Stat. unc. ($\rm pb$) & $(1+\delta)$ & Dressed ($\rm pb$) & $|1-\Pi(s)|^{-2}$ & Born $\pm$ Stat. $\pm$ Sys. ($\rm pb$) & Rel. Sys. (\%)\\
\hline
\endhead 
\hline
\endfoot 
\endlastfoot


 3.8077           & 238            & 26             & 5.5293         & 43             & 1.0562         & 41              $\pm$     4 $\pm$ 2               & 4.85       \\  
3.8480           & 175            & 67             & 1.1286         & 155            & 1.0521         & 147             $\pm$    57 $\pm$ 5               & 3.43       \\  
3.8880           & 1171           & 113            & 0.7968         & 1469           & 1.0494         & 1400            $\pm$   135 $\pm$ 59              & 4.24       \\  
3.8930           & 1200           & 121            & 0.7916         & 1516           & 1.0489         & 1446            $\pm$   145 $\pm$ 60              & 4.17       \\  
3.8962           & 1299           & 42             & 0.7870         & 1650           & 1.0487         & 1574            $\pm$    51 $\pm$ 46              & 2.95       \\  
3.8980           & 1394           & 112            & 0.7849         & 1776           & 1.0485         & 1694            $\pm$   136 $\pm$ 86              & 5.08       \\  
3.9030           & 1590           & 122            & 0.7808         & 2036           & 1.0483         & 1942            $\pm$   149 $\pm$ 116             & 5.96       \\  
3.9080           & 1713           & 121            & 0.7796         & 2198           & 1.0481         & 2097            $\pm$   148 $\pm$ 83              & 3.94       \\
\end{longtable}

\clearpage
\begin{longtable}[h]{c  c  c  c c  c c c c}
\caption{Summary of $\sigma(e^+e^- \to D^+X)$. The ISR factor values are calculated directly from the formula rather than from the \textsc{hybrid} generator simulation. (150 points)} 	\\
\hline
\centering
$\sqrt{s}$ (GeV) & Obs. ($\rm pb$) & Stat. unc. ($\rm pb$) & $(1+\delta)$ & Dressed ($\rm pb$) & $|1-\Pi(s)|^{-2}$ & Born $\pm$ Stat. $\pm$ Sys. ($\rm pb$) & Rel. Sys. (\%)\\
\hline
\endfirsthead 
\hline      
$\sqrt{s}$ (GeV) & Obs. ($\rm pb$) & Stat. unc. ($\rm pb$) & $(1+\delta)$ & Dressed ($\rm pb$) & $|1-\Pi(s)|^{-2}$ & Born $\pm$ Stat. $\pm$ Sys. ($\rm pb$) & Rel. Sys. (\%)\\
\hline
\endhead 
\hline
\endfoot 
\endlastfoot


3.8077           & 183            & 14             & 5.6940         & 32             & 1.0562         & 30              $\pm$     2 $\pm$ 1               & 4.42       \\  
3.8480           & 225            & 35             & 1.0310         & 218            & 1.0521         & 207             $\pm$    32 $\pm$ 12              & 5.65       \\  
3.8880           & 517            & 53             & 0.8142         & 634            & 1.0494         & 605             $\pm$    62 $\pm$ 36              & 5.91       \\  
3.8930           & 734            & 56             & 0.8048         & 912            & 1.0489         & 869             $\pm$    66 $\pm$ 49              & 5.59       \\  
3.8962           & 688            & 20             & 0.7974         & 862            & 1.0487         & 822             $\pm$    23 $\pm$ 32              & 3.84       \\  
3.8980           & 749            & 52             & 0.7926         & 945            & 1.0485         & 902             $\pm$    63 $\pm$ 25              & 2.76       \\  
3.9030           & 856            & 54             & 0.7859         & 1090           & 1.0483         & 1040            $\pm$    65 $\pm$ 24              & 2.36       \\  
3.9080           & 994            & 61             & 0.7835         & 1269           & 1.0481         & 1211            $\pm$    74 $\pm$ 35              & 2.85       \\
\end{longtable}

\clearpage
\begin{longtable}[h]{c  c  c  c c  c c c c}
\caption{Summary of $\sigma(e^+e^- \to D^+_sX)$. The ISR factor values are calculated directly from the formula rather than from the \textsc{hybrid} generator simulation. (150 points)} 	\\
\hline
\centering
$\sqrt{s}$ (GeV) & Obs. ($\rm pb$) & Stat. unc. ($\rm pb$) & $(1+\delta)$ & Dressed ($\rm pb$) & $|1-\Pi(s)|^{-2}$ & Born $\pm$ Stat. $\pm$ Sys. ($\rm pb$) & Rel. Sys. (\%)\\
\hline
\endfirsthead 
\hline
$\sqrt{s}$ (GeV) & Obs. ($\rm pb$) & Stat. unc. ($\rm pb$) & $(1+\delta)$ & Dressed ($\rm pb$) & $|1-\Pi(s)|^{-2}$ & Born $\pm$ Stat. $\pm$ Sys. ($\rm pb$) & Rel. Sys. (\%)\\
\hline
\endhead 
\hline
\endfoot
\endlastfoot


3.8077            & --             & --             & --             & --             &--              & --                               & --       \\
3.8480            & --             & --             & --             & --             &--              & --                               & --       \\
3.8880            & --             & --             & --             & --             &--              & --                               & --       \\
3.8930            & --             & --             & --             & --             &--              & --                               & --       \\
3.8962            & --             & --             & --             & --             &--              & --                               & --       \\
3.8980            & --             & --             & --             & --             &--              & --                               & --       \\
3.9030            & --             & --             & --             & --             &--              & --                               & --       \\
3.9080            & --             & --             & --             & --             &--              & --                               & --       \\
\end{longtable}

\clearpage
\begin{longtable}[h]{c  c | c c  | c c | c c}
\caption{Summary of the measured $R_D$ values. (150 points)} 	\\
\hline
$\sqrt{s}$ (GeV) &    $R_D$&     $\sqrt{s}$ (GeV) &    $R_D$&  $\sqrt{s}$ (GeV) &    $R_D$  &$\sqrt{s}$ (GeV) &    $R_D$\\
\hline
\endfirsthead 
\hline
 $\sqrt{s}$ (GeV) &    $R_D$&     $\sqrt{s}$ (GeV) &    $R_D$&  $\sqrt{s}$ (GeV) &    $R_D$ & $\sqrt{s}$ (GeV) &    $R_D$ \\
\hline
\endhead 
\hline
\endfoot 
\endlastfoot

3.8077     &  0.01 $\pm$ 0.00       &  4.0480     &  2.29 $\pm$ 0.08      & 4.2330     &  0.80 $\pm$ 0.05   &   4.3964     &  1.76 $\pm$ 0.04  \\
3.8480     &  0.06 $\pm$ 0.01       &  4.0530     &  2.13 $\pm$ 0.08      & 4.2357     &  0.77 $\pm$ 0.02   &   4.3980     &  1.83 $\pm$ 0.08  \\
3.8880     &  0.35 $\pm$ 0.03       &  4.0580     &  2.07 $\pm$ 0.08      & 4.2380     &  0.88 $\pm$ 0.04   &   4.4080     &  2.08 $\pm$ 0.11  \\
3.8930     &  0.41 $\pm$ 0.04       &  4.0630     &  2.12 $\pm$ 0.08      & 4.2410     &  0.74 $\pm$ 0.04   &   4.4156     &  1.82 $\pm$ 0.04  \\
3.8962     &  0.42 $\pm$ 0.01       &  4.0680     &  1.99 $\pm$ 0.08      & 4.2417     &  0.71 $\pm$ 0.02   &   4.4180     &  1.89 $\pm$ 0.07  \\
3.8980     &  0.45 $\pm$ 0.03       &  4.0780     &  1.98 $\pm$ 0.07      & 4.2430     &  0.72 $\pm$ 0.04   &   4.4230     &  1.80 $\pm$ 0.07  \\
3.9030     &  0.52 $\pm$ 0.04       &  4.0854     &  1.86 $\pm$ 0.06      & 4.2438     &  0.71 $\pm$ 0.02   &   4.4280     &  2.04 $\pm$ 0.08  \\
3.9080     &  0.58 $\pm$ 0.03       &  4.0880     &  1.98 $\pm$ 0.09      & 4.2460     &  0.78 $\pm$ 0.04   &   4.4362     &  1.85 $\pm$ 0.04  \\
\end{longtable}

\clearpage

\begin{longtable}[h]{c  c   |    c  c | c c | c c}
\caption{Summary of integrated luminosities for  data samples. (150 points)} 	\\
\hline
  $\sqrt{s}$ (GeV) &    $\mathcal{L}$ ($\rm pb^{-1}$) &  $\sqrt{s}$ (GeV) &    $\mathcal{L}$ ($\rm pb^{-1}$) &  $\sqrt{s}$ (GeV) &    $\mathcal{L}$ ($\rm pb^{-1}$) &  $\sqrt{s}$ (GeV) &    $\mathcal{L}$ ($\rm pb^{-1}$)\\
\hline
\endfirsthead 
\hline
  $\sqrt{s}$ (GeV) &    $\mathcal{L}$ ($\rm pb^{-1}$) &  $\sqrt{s}$ (GeV) &    $\mathcal{L}$ ($\rm pb^{-1}$) &  $\sqrt{s}$ (GeV) &    $\mathcal{L}$ ($\rm pb^{-1}$) &  $\sqrt{s}$ (GeV) &    $\mathcal{L}$ ($\rm pb^{-1}$)\\
\hline
\endhead 
\hline
\endfoot 
\endlastfoot 
3.8077  &  50.54      & 4.0480  &  6.57      &     4.2330  &  8.27      &      4.3964  &  507.80     \\  
3.8480  &  7.97       & 4.0530  &  6.93      &     4.2357  &  530.30    &      4.3980  &  7.18       \\
3.8880  &  7.76       & 4.0580  &  6.34      &     4.2380  &  7.83      &      4.4080  &  6.35       \\
3.8930  &  7.57       & 4.0630  &  7.02      &     4.2410  &  8.57      &      4.4156  &  1090.70    \\   
3.8962  &  52.61      & 4.0680  &  7.27      &     4.2417  &  55.88     &      4.4180  &  7.52       \\
3.8980  &  7.58       & 4.0780  &  7.72      &     4.2430  &  8.49      &      4.4230  &  7.44       \\
3.9030  &  7.60       & 4.0854  &  52.86     &     4.2438  &  538.10    &      4.4280  &  6.79       \\
3.9080  &  7.24       & 4.0880  &  7.61      &     4.2460  &  8.55      &      4.4362  &  569.90     \\
\end{longtable}


